\documentclass[pra,10pt,twocolumn,superscriptaddress,showpacs]{revtex4-1}
\usepackage{amsmath,amsthm}
\usepackage{latexsym}
\usepackage{amssymb}
\usepackage{mathtools}
\usepackage{braket}
\usepackage{graphicx}
\usepackage{dsfont}
\usepackage[colorlinks=true, citecolor=blue, urlcolor=blue]{hyperref}
\usepackage{amsfonts}
\makeatletter
\newtheorem*{rep@theorem}{\rep@title}
\newcommand{\newreptheorem}[2]{%
\newenvironment{rep#1}[1]{%
 \def\rep@title{#2 \ref{##1}}%
 \begin{rep@theorem}}%
 {\end{rep@theorem}}}
\makeatother

\newreptheorem{theorem}{Theorem}

\begin{document}
\title{Monogamy of entanglement of maximal dimension}
\author{Sumit Nandi}
\email{sumit.enandi@gmail.com}
\affiliation{Purandarpur High School, Purandarpur, West Bengal, 731129, India}

%

\begin{abstract}
In the present paper, a trade off of sharing of entanglement between subsystems of a higher dimensional quantum state is derived. It is presented in terms of an inequality which is analogous to the Coffman-Kundu-Wootters inequality that succinctly describes monogamy of entanglement in $\mathcal{C}^2\otimes \mathcal{C}^2\otimes \mathcal{C}^2$ dimensional pure state. To derive the monogamy inequality in $\mathcal{C}^d\otimes \mathcal{C}^d\otimes \mathcal{C}^d$ dimension, G-concurrence measure of entanglement is considered as a measure of entanglement of maximal dimension. The approach of the present paper incidentally points towards a rigorous framework which enables us to obtain an upper bound of G-concurrence of a bipartite qudit mixed state. Obtained upper bound of G-concurrence is then shown to satisfy a monogamy relation.  
\end{abstract}
\maketitle
\section{Introduction}

Entanglement manifests several counter-intuitive phenomenon in quantum mechanics making it remarkably distinctive from its classical counterpart. Unlike classical correlation, entanglement cannot be shared freely among its different subsystems. Nature imposes severe restriction on sharing entanglement arbitrarily between the subsystems of a given multipartite state, and this
is  referred to as \emph{monogamy of entanglement} \cite{ckw}. For example, given a tripartite state shared by three parties namely, A, B and C, if A and B are maximally entangled between themselves,  there cannot be any entanglement between A (B) and C. 
It had been shown that monogamy of entanglement plays an instrumental role in establishing security in the entanglement based quantum key distribution protocols \cite{qkd1, qkd2}. The importance of monogamy of quantum correlation was harnessed in the context of some foundational issues - for example, it was shown in \cite{acin} that quantum cloning and state estimation are equivalent in the asymptotic regime \cite{acin}. Monogamy of quantum correlations is also found to be important in other branches of physics such as condensed-matter physics \cite{ma}, frustrated
spin systems \cite{rao}, statistical physics \cite{ben}, black-hole physics \cite{loyd} \emph{etc}.\\~\\
The notion of monogamy of entanglement was first formulated in a form of inequality known as the Coffman-Kundu-Wootters (CKW) inequality \cite{ckw}. The narrative nicely  demonstrates a trade-off between shared entanglement of A with B and C. To derive the trade-off relation in $\mathcal{C}^2\otimes \mathcal{C}^2\otimes \mathcal{C}^2$, concurrence \cite{con1,con2} was used as a measure of entanglement. Then, it was realised that monogamy is valid for a large system and generalised for arbitrary multi-qubit states in \cite{osborne}. Remarkably, the monogamous aspect of entanglement was also
revealed by other measures of entanglement. Monogamous nature of squashed entanglement and Tsallis-q entanglement were described in \cite{sq} and \cite{q}, respectively. 
It has also been shown that squared entanglement of formation satisfies monogamy relation in multiqubit mixed states
\cite{mixed}. A monogamy inequality was presented for a specific class of multi-qudit systems by convex-roof extended negativity \cite{kim_das}.
The authors in \cite{li1} defined a qudit measure of entanglement to study monogamy relation in tripartite qutrit system. Recently, monogamy of entanglement was shown for generalized $n$-qudit W-class states \cite{alpha} taking R\'{e}nyi-$\alpha$ entropy measure of 
entanglement. However, a general monogamy relation as such CKW inequality for multipartite qudit systems has not been composed till now. It was observed that CKW inequality does no longer hold in higher dimension - a counter example was provided in \cite{yong} to show that the CKW inequality does not hold for a tripartite qutrit state. Notably, CKW inequality was derived using concurrence \cite{con1} measure of entanglement. Thus generalization of CKW inequality might require a suitable framework to quantify entanglement of higher dimensional systems.
Nonetheless, a CKW type inequality for qudit systems was conjectured in \cite{gour3}. The author therein used concurrence monotones to quantify $d$-level entanglement.
\\
 In the context of pure bipartite states, Schmidt coefficients are known to play a pivotal role towards quantification of qudit entanglement. In order to characterise entanglement of a pure bipartite qudit state, a finite set of functions of the Schmidt coefficients known as entanglement monotones was constructed  in \cite{vidal}. Subsequently, in \cite{gour2}, another set of functions of the Schmidt coefficients, namely, concurrence monotones was introduced. The later set of monotones consists of $(d-1)$ non-trivial functions of the Schmidt coefficients. The last monotone of this family is known as G-concurrence that measures to which extent the maximum Schmidt rank (Schmidt number for mixed states, see \cite{sanpera}) is contained in a given qudit joint state. Although, G-concurrence is easily computable for pure states, the mixed state generalisation is not achieved till date.
In the present context, I shall address the issue by developing a framework to obtain an upper bound of G-concurrence of mixed qudit bipartite states. Then, a detailed analytical proof of a CKW type inequality in higher dimension conjectured in \cite{gour3} will be provided. I will also illustrate the validity of the derived inequality by providing examples of few important classes of genuine entangled states in $\mathcal{C}^3\otimes \mathcal{C}^3\otimes \mathcal{C}^3$, and $\mathcal{C}^d\otimes \mathcal{C}^d\otimes \mathcal{C}^d$ respectively. 
\\~\\
The paper is organised as follows. In Sec. (\ref{preli}) we  briefly recapitulate G-concurrence measure of entanglement. In Sec. (\ref{main}) the main result is presented in the form of a theorem. The validity of derived monogamy inequality is described by illuminating examples of some important classes of states in Sec. (\ref{example}). Finally, we note down some concluding remarks in Sec. (\ref{discuss}).

\section{Concurrence monotones and G-concurrence }\label{preli}

Concurrence monotones \cite{gour2} are presented to characterise entanglement of a pure bipartite qudit state. These monotones are obtained by constructing symmetric functions of Schmidt coefficients of the given state. Each of these monotones plays significant role to study $d$-level non-local aspects of a pure state. For a pure state $\ket{\psi}$ in $\mathcal{C}^d\otimes \mathcal{C}^d$ with the  Schmidt coefficients $\lambda_0$, $\dots$, $\lambda_{d-1}$ respectively, one can define its $d$ concurrence monotones \cite{gour2} in the following way:
\begin{eqnarray}
C_1(\ket{\psi})&=&\sum_{i=0}^{d-1}\lambda_i^2\nonumber\\ C_2(\ket{\psi})&=&d\big(\sum_{i<j}^{d-1}\lambda_i^2
\lambda_j^2
\big)^{\frac{1}{2}} \nonumber\\
.\nonumber\\
.\nonumber\\
C_d(\ket{\psi})&=&d\big(\prod_{i=0}^{d-1} \lambda_i^2\big)^{\frac{1}{d}} \label{gcon}.
\end{eqnarray} 
The first monotone is trivial, and sums up to $one$ due to the normalization constraint. $C_2$ can be readily readily recognised as concurrence \cite{con1,con2} in usual sense. Amongst the others, the last monotone, known as G-concurrence, is of particular interest in our context. Henceforth, we shall denote it as $G(\ket{\psi})$ for our own convenience. For a $2\otimes 2$ pure state $\ket{\psi}$, it agrees with the concurrence presented in \cite{con1}. The quantity G-concurrence can be computed by evaluating the determinant of one of the reduced density matrices of a given pure state. G-concurrence can also be obtained for mixed states by convex roof extension. Given a mixed state $\rho\in \mathcal{C}^d\otimes C^d$, G-concurrence of $\rho$ is the minimized average G-concurrence of an ensemble of pure states $\ket{\psi_i}$:
\begin{equation}\label{gcon_mixed_state}
G(\rho)=\text{inf}_{\{\ket{\psi_j}\}}\sum_{i}G(|\psi_i\rangle),
\end{equation}
where the infimum is taken over all pure state decompositions realising $\rho=\sum_j|\psi_j\rangle\langle\psi_j|$ and $G(|\psi_j\rangle)$ is the G-concurrence of $\ket{\psi_j}$ which can be determined using Eq.(\ref{gcon}). However, we note that a closed expression of Eq.(\ref{gcon_mixed_state}) has not been 
obtained, and remains broadly unexplored. In the next, we will alleviate it by formulating a suitable framework for computing an upper bound of G-concurrence of an arbitrary mixed state in $\mathcal{C}^d\otimes \mathcal{C}^d$.


\section{Proof of monogamy inequality}\label{main}

Let us consider a tripartite state $\ket{\Psi}_{123}$ of Hilbert space dimension $\mathcal{C}^d\otimes \mathcal{C}^d \otimes \mathcal{C}^d$ $(d> 2)$ which is shared between three observers $1$, $2$ and $3$. We denote the bipartitions consisting of the subsystems $1, 2$ as $\rho_{12}$, $1, 3$ as $\rho_{13}$, and the bipartition across $1:23$ as $\rho_{1(23)}$, respectively. We then proceed to prove a conjecture provided in \cite{gour3} which describes an entanglement trade-off quantified by G-concurrence of  $\rho_{12}$, $\rho_{13}$ and $\rho_{1(23)}$, respectively. \\~\\
\textbf{Theorem 1} For any pure state $\ket{\Psi}_{123}$, the entanglement quantified by G-concurrence of $\rho_{12}$, $\rho_{13}$ and $\rho_{1(23)}$ denoted as $G_{12}$, $G_{13}$ and $G_{1(23)}$ satisfy the following monogamy inequality
\begin{equation}\label{ineq}
G^d_{12}+G^d_{13}\le G^d_{1(23)},
\end{equation}
\\~\\
\textbf{Proof:} Let $\{\ket{i}\}$, $\{\ket{j}\}$ and $\{\ket{k}\}$ denote an orthonormal basis spanning the Hilbert space $\mathcal{C}^d\otimes \mathcal{C}^d \otimes \mathcal{C}^d$, and we write the state $\ket{\Psi}_{123}\in \mathcal{C}^d\otimes \mathcal{C}^d \otimes \mathcal{C}^d$, $(d> 2)$   
\begin{equation}\label{eq4}
\ket{\Psi}_{123}=\sum_{i,j,k=0}^{d-1}a_{ijk}\ket{ijk},
\end{equation} 
where $a_{ijk}\in\mathbb{C}$ and satisfies the normalization constraint $\sum_{i,j,k}|a_{ijk}|^2=1$. 
We obtain the subsystem $\rho_{12}$ by tracing over the third subsystem $3$,
\begin{equation}
\rho_{12}=\sum_{i,i^\prime,j,j^\prime,k}a^*_{i^\prime j^\prime k}a_{ijk}|i^\prime j^\prime\rangle\langle ij|.
\end{equation}
Although the Hilbert space describing $\rho_{12}$ is of  $d^2$ dimensional, it would be sufficient to consider only $d$ of them which are necessary to be entangled with the single qudit subsystem of the observer $3$ \cite{ckw}. Henceforth, we will consider the rank $r$ of $\rho_{12}$ as $r(\rho_{12})=d$, and the same for the other subsystem $\rho_{13}$ shared by the observers $1$ and $3$. Now, the mixed state $\rho_{12}$ can be realized by the following pure state decomposition 
\begin{equation}\label{subsystem12}
\rho_{12}=\sum_{k=0}^{d-1}|\phi_k\rangle\langle\phi_k|,
\end{equation} 
where $\ket{\phi_k}=\sum_{ij}a_{ijk}\ket{ij}$ are sub-normalized orthogonal state vectors with norm no greater than $one$. The range of $\rho_{12}$ is spanned by these three vectors. A closed expression of G-concurrence, often denoted as $G({\rho_{ij}})$, of a general qudit mixed state $\rho_{ij}$ does not exist. We circumvent the situation by developing a framework to obtain an upper bound of $G({\rho_{ij}})$. To achieve our goal, we invoke the lemma presented in \cite{gour4}  which states that $\rho_{12}$ must contain one such pure state decomposition  $\{\ket{\xi_m}\}$ in which G-concurrence is non-vanishing for a \emph{particular} pure state in that decomposition, and vanishes for all other pure states in that decomposition. We consider the pure state decomposition of $\rho_{12}$,  
\begin{equation}\label{decomp2}
\rho_{12}=\sum_{l=0}^{d^\prime}|\chi_l\rangle\langle\chi_l|
\end{equation} 
which is obtained using the $d^\prime \times d$ isometry $\mathcal{U}=(U_{lk})$ i.e. $\mathcal{U}^\dagger \mathcal{U}=\mathbb{I}_{d}$, such that 
\begin{equation}\label{isometry}
\ket{\chi_l}=\sum_kU_{lk}\ket{\phi_k}.
\end{equation}
To evaluate G-concurrence of $\rho_{12}$ realising pure state decomposition $\{\ket{\chi_l}\}$, without loss of generality, we assume 
\begin{equation}\label{condition}
G(\ket{\chi_0})\neq 0 \hspace{.2in} \text{and} \hspace{.2in} G(\ket{\chi_l})=0, \hspace{.051in}\forall\hspace{.051in} l=1\dots (d^{\prime}-1)
\end{equation}  
Thus one can obtain G-concurrence of the mixed state $\rho_{12}$ by calculating only G-concurrence of the pure state $\ket{\chi_0}$ in the ensemble, and hence, we write 
\begin{equation}\label{eq10}
G_{12}=G(\rho_{12})=G(|\chi_0\rangle\langle \chi_0|).
\end{equation}

Before we proceed further, we next show G-concurrence yields minimum value in this particular pure state decomposition $\{\ket{\chi_s}\}$ with the mentioned property stated in Eq.(\ref{condition}). To prove that, let us consider the following pure state decomposition $\rho_{12}^\ast$
\begin{equation}\label{new decomp}
\rho_{12}^\ast=\sum_l|\nu_l\rangle\langle \nu_l|
\end{equation}
where $\ket{\nu_l}=\sum_j W_{lj}\ket{\chi_j}$. So, G-concurrence of $\rho_{12}^\ast$ is given by 
\begin{eqnarray}
G(\rho_{12}^\ast)&=&G\big(\sum_{l,j,j^\prime}
W^\ast_{lj^\prime}W_{lj}|\chi_j\rangle\langle \chi_j^\prime|\big)\\
&\ge&  \sum_{l,j,j^\prime}
W^\ast_{lj^\prime}W_{lj}G(|\chi_j\rangle\langle \chi_j^\prime|)\\
&\ge& \sum_l W^\ast_{l0}W_{l0}G(|\chi_0\rangle\langle \chi_0|)\\
&=& G(|\chi_0\rangle\langle \chi_0|)\label{Eq.15} \\
&=&G(\rho_{12})
\end{eqnarray}
We use concavity of G-concurrence \cite{gour2} at intermediate stage, and invoke unitary properties of the elements $W_{lj}$ before Eq.(\ref{Eq.15}), respectively.
It suffices now to compute G-concurrence of the particular pure state $\ket{\chi_0}$ to obtain G-concurrence of $\rho_{12}$.

We recall Eq.(\ref{isometry}), and write
\begin{eqnarray}
\ket{\chi_l}&=&\sum_k U_{lk}\ket{\phi_k}\\
&=&\sum_k\sum_{i,j} U_{lk}a_{ijk}\ket{ij}\\
&=&\sum_{i,j} \tilde{a}_{ijl}\ket{ij}
\end{eqnarray}
where $\tilde{a}_{ijl}=\sum_{k}U_{lk}a_{ijk}$.
Let us now evaluate G-concurrence of $\ket{\chi_0}$ by computing its Schmidt coefficients. From the above equation  it readily follows
\begin{equation}
\ket{\chi_0}=\sum_{i,j} \tilde{a}_{ij0}\ket{ij}
\end{equation}

 Substituting $\sum_{j}\tilde{a}_{ij0}\ket{j}=\lambda_{i0}\ket{\tilde{i}}$ into the above expression, we obtain  
\begin{eqnarray}
\ket{\chi_0}=(X_0\otimes \mathbb{I})\sum_i\ket{i \tilde{i}}=\sum_i\lambda_{i0}\ket{i\tilde i},
\end{eqnarray}   
where $X_0$ is a $d\times d$ diagonal matrix with the entries $\lambda_{i0}$ respectively. Therefore, G-concurrence of the unnormalised vector $\ket{\chi_0}$ is \cite{gour4}
\begin{equation}
G(\ket{\chi_0})=\big(\prod_i|\lambda_{i0}|^2\big)^{\frac{1}{d}},
\end{equation}
$d^{th}$ power of which is given by
\begin{equation}\label{eq15}
G^d(\ket{\chi_0})=\prod_i|\lambda_{i0}|^2.
\end{equation}
Thus, $G^d_{12}$ can be recast in a suggestive way as given below  
\begin{eqnarray}
G^d_{12}&=&|\lambda_{00}|^2|\lambda_{10}|^2 \dots|\lambda_{(d-1)0}|^2
\end{eqnarray}
To express the Schmidt coefficients $\lambda$'s in terms of $a_{ijk}$, we write:
\begin{eqnarray}
|\lambda_{i0}|^2 = \sum_{j,j^\prime}\tilde{a}_{ij^\prime 0}^*\tilde{a}_{ij0}
\langle j^\prime|j\rangle =
\sum_{j}|\tilde{a}_{ij0}|^2
\end{eqnarray}
where we have used the fact that $\{\ket{j}\}$ constitutes an orthonormal basis.
So, one can rewrite the above equation
\begin{eqnarray}
|\lambda_{i0}|^2 &=&\sum_j|\tilde{a}_{0j0}|^2\sum_j|\tilde{a}_{1j0}|^2\dots \sum_j|\tilde{a}_{(d-1)j0}|^2
\\ &=&\sum_j|\sum_{k} U_{0k}a_{0jk}|^2\sum_j|\sum_{k}U_{0k}a_{1jk}|^2\dots \nonumber\\ && \sum_j|\sum_{k}U_{0k}a_{(d-1)jk}|^2
\end{eqnarray}
Using the Cauchy-Schwarz inequality, we obtain
\begin{eqnarray}\label{g12}
G^d_{12}&\le & \sum_{k}|U_{0k}|^2\sum_{j,k}|a_{0jk}|^2\sum_{k}|U_{0k}|^2\sum_{j,k}|a_{1jk}|^2\dots \nonumber\\&&
\sum_{k}|U_{0k}|^2\sum_{j,k}|a_{(d-1)jk}|^2\nonumber\\
&\le &  \sum_{j,k}|a_{0jk}|^2\sum_{j,k}|a_{1jk}|^2\dots\sum_{j,k}|a_{(d-1)jk}|^2
 \nonumber\\
\end{eqnarray}
Where we have used the unitary property $\sum_{k}|U_{0k}|^2=1$.\\
In a similar approach, G-concurrence of the other subsystem $\rho_{13}$ can be derived. We obtain a similar expression as given by Eq.(\ref{g12}) for $G^d_{13}$. So, we can write down the sum as  
\begin{equation}\label{sum}
G^d_{12}+G^d_{13}\le 2\sum_{j,k}|a_{0jk}|^2\sum_{j,k}|a_{1jk}|^2 \dots \sum_{j,k}|a_{(d-1)jk}|^2
\end{equation}
 It is to be noted that our theorem (\ref{ineq}) aims to provide an algebraic constraint between $G_{12}$, $G_{13}$ and $G_{1(23)}$. In order to achieve that specific goal, it remains left to evaluate $G_{1(23)}$, G-concurrence of the subsystem across $1|23$. So, we write Eq.(\ref{eq4}) as
\begin{eqnarray}\label{schmidt_form}
\ket{\psi}_{1(23)}&=&\sum_{i,j,k=0}^{d-1}a_{ijk}\ket{ijk}\nonumber\\
&=&\sum_i\Gamma_{i}\ket{i}\ket{\zeta_i},
\end{eqnarray} 
 where $\Gamma_i\ket{{\zeta_i}}=\sum_{j,k}a_{ijk}\ket{jk}$. Writing this way, it presumes $\ket{\psi}_{123}$ as a bipartite state $\ket{\psi}_{1(23)}$ shared between the subsystem $1$ and the subsystem $23$ composed of the subsystem $2$ and $3$ respectively. Schmidt coefficients of  $\ket{\psi}_{1(23)}$ are given by $|\Gamma_i|$. So, we obtain
\begin{equation}
|\Gamma_i|^2=\sum_{j,k}|a_{ijk}|^2
\end{equation} 
Thus, one can compute G-concurrence of the pure bipartite state $\ket{\psi}_{1(23)}$ $\in \mathcal{C}^d\otimes\mathcal{C}^{2d}$ using the prescription given in \cite{gour2}. Therefore,
\begin{eqnarray}\label{gcon_rho_1-23}
G^d_{1(23)}&=&d^d|\Gamma_0|^2|\Gamma_1|^2\dots |\Gamma_{(d-1)}|^2\\
&=&d^d\sum_{j,k}|a_{0jk}|^2\sum_{j,k}|a_{1jk}|^2\dots \sum_{j,k}|a_{(d-1)jk}|^2\label{rho1}\nonumber
\end{eqnarray}
Now comparing it with Eq. (\ref{sum}), we obtain 
\begin{eqnarray}\label{gcon_ineq}
G^d_{12}+G^d_{13}\le G^d_{1(23)}\label{proof}
\end{eqnarray}
This completes the proof of our theorem (\ref{ineq}). In Appendix \ref{appendix1}, we construct an explicit proof for $d=3$.


\section{Examples}\label{example}
In this section we will present few illustrating examples to show the validity of our theorem for some important classes of states. We start with the tripartite qutrit state introduced in \cite{yong} 
\begin{equation}
\ket{\chi}=\frac{1}{\sqrt6}(\ket{012}-\ket{021}+\ket{120}-\ket{102}+\ket{201}-\ket{210})
\end{equation} 
For the above state it can be seen that
\begin{eqnarray}
\rho_{12}&=&\frac{1}{3}(|x\rangle\langle x|+|y\rangle\langle y|+|z\rangle\langle z|)\\
\rho_{13}&=&\frac{1}{3}(|x\rangle\langle x|+|y\rangle\langle y|+|z\rangle\langle z|)\\
\psi_{1(23)}&=&\frac{1}{\sqrt{3}}\big(\ket{0}\ket{\mu_0}+\ket{1}\ket{\mu_1}+
\ket{2}\ket{\mu_2}\big)
\end{eqnarray}
where 
\begin{eqnarray}
\ket{x}=\frac{1}{\sqrt 2}(\ket{01}-\ket{10})\\
\ket{y}=\frac{1}{\sqrt 2}(\ket{02}-\ket{20})\\
\ket{z}=\frac{1}{\sqrt 2}(\ket{12}-\ket{21})\\
\ket{\mu_0}=\frac{1}{\sqrt2 }(\ket{12}-\ket{21})\\
\ket{\mu_1}=\frac{1}{\sqrt2 }(\ket{20}-\ket{02})\\
\ket{\mu_2}=\frac{1}{\sqrt2 }(\ket{01}-\ket{10})
\end{eqnarray}
The Schmidt coefficients of $\ket{x}$, $\ket{y}$ and $\ket{z}$ are $\frac{1}{\sqrt 2}$, $\frac{1}{\sqrt 2}$ and $0$ respectively. So, G-concurrence for these states vanish, and hence, $G_{12}=0=G_{13}$. To evaluate $G_{1(23)}$, we find the Schmidt coefficients of $\psi_{1(23)}$ to be $\{\frac{1}{\sqrt 3},\frac{1}{\sqrt 3},\frac{1}{\sqrt 3}\}$. Thus, $G_{1(23)}=1$, and immediately it follows 
that $G^3_{1(23)}>G^3_{12}+G^3_{13}$. 
\\
Next,  we consider the tripartite generalized W-class state in $\mathcal{C}^3\otimes \mathcal{C}^3\otimes \mathcal{C}^3$ as follows
\begin{eqnarray}\label{wstate}
\ket{\mathcal{W}}&=&a_{11}\ket{100}+a_{12}\ket{200}+
a_{21}\ket{010}+\nonumber\\&&a_{22}\ket{020}+a_{31}\ket{001}+
a_{32}\ket{002},
\end{eqnarray}
where $|a_{11}|^2+|a_{12}|^2+|a_{21}|^2+|a_{22}|^2+|a_{31}|^2+|a_{32}|^2=1$. It is noted if all the elements $a_{ij}$ assume $\frac{1}{\sqrt 6}$, one obtains the familiar W-state. The subsystem $\rho_{12}$ is obtained by tracing over the third subsystem, and given by
\begin{equation}
\rho_{12}=\ket{\tilde{x}}\langle \tilde{x}|+\ket{\tilde{y}}\langle \tilde{y}|,
\end{equation}
where we write the un-normalized kets $\ket{\tilde{x}}$ and $\ket{\tilde{y}}$ as
\begin{eqnarray}
\ket{\tilde{x}} &=& a_{11}\ket{10}+a_{12}\ket{20}+a_{21}
\ket{01}+a_{22}\ket{02}\\
\ket{\tilde{y}} &=&\sqrt{|a_{31}|^2+|a_{32}|^2}\ket{00}.
\end{eqnarray}
At this point we shall follow the steps \cite{1hjw} to evaluate the G-concurrence of the mixed state $\rho_{12}$. One can find another pure state decomposition of $\rho_{12}$ by suitable unitary transformation. We write one such pure state decomposition as 
$\rho_{12}$$=$$\ket{\tilde{\phi}_1}\langle \tilde{\phi_1}|+\ket{\tilde{\phi}_2}\langle \tilde{\phi_2}|$, where 
\begin{eqnarray}
\ket{\tilde{\phi}_1}&=&u_{11}\ket{x}+u_{12}\ket{y}\\
\ket{\tilde{\phi}_2}&=&u_{21}\ket{x}+u_{22}\ket{y}.
\end{eqnarray}
Here $u_{hl}$ are elements of the $r\times r$ unitary matrix, $r$ being the rank of $\rho_{12}$. For this particular case, $r(\rho_{12})=2$. 
Thus we rewrite $\rho_{12}$ as 
\begin{eqnarray}
\rho_{12}&=&p_1\ket{\phi_1}\langle \phi_1|+p_2\ket{\phi_2}\langle \phi_2|\\
&=&\sum_{h=1}^rp_h\ket{\phi_h}\langle \phi_h|,
\label{decomposition}
\end{eqnarray}
where we have defined $p_i=\langle\tilde{\phi_i}|\tilde{\phi_i}\rangle$ and $\ket{\phi_i}=\frac{\ket{\tilde{\phi_i}}}{\sqrt{p_i}}$. Now the G-concurrence of $\rho_{12}$ is 
\begin{eqnarray}
G_{12}&=&{min}\sum_{h=1}^rp_hG(\ket{\phi_h})\nonumber\\
&=&p_1 G(\ket{\phi_1})+p_2G(\ket{\phi_2})\nonumber\\
&=&p_1G(\frac{\ket{\tilde{\phi_1}}}{\sqrt{p_1}})+p_2 G(\frac{\ket{\tilde{\phi_2}}}{\sqrt{p_2}})
\end{eqnarray}
It can be verified  that $G(\frac{\ket{\tilde{\phi_i}}}{\sqrt{p_i}})$
vanishes for $i=1$, $2$. So we obtain
\begin{equation}
G_{12}=0
\end{equation}
Since, the G-concurrence vanishes for this particular decomposition, it represents the infimum pure state decomposition. In a similar fashion, one can obtain  $G_{13}=0$. Next, to  evaluate $G_{1(23)}$, we rewrite Eq. (\ref{wstate}) as follows
\begin{equation}
\ket{\mathcal{W}}=\frac{1}{\sqrt 2}(\ket{\tilde{+}}\ket{00}+\ket{0}\ket{\tilde{00}})
\end{equation}
where $\ket{\tilde{+}}=\sqrt{2}(a_{11}\ket{1}+a_{12}\ket{2})$ and $\ket{\tilde{00}}=\sqrt{2}(a_{21}\ket{10}+a_{22}\ket{20}+a_{31}\ket{01}+a_{32}\ket{02})$.
It is to be noted that $\langle{\tilde{+}}|0\rangle=0$ and $\langle{\tilde{00}}|00\rangle=0$. The Schmidt coefficients are $\{\frac{1}{\sqrt 2},\frac{1}{\sqrt 2},0\}$, and $G_{1(23)}$ vanishes. It shows that the $\ket{\mathcal{W}}$ state saturates our monogamy inequality,
\begin{equation} 
G^3_{1(23)}=G^3_{12}+G^3_{13}
\end{equation}  

Lastly, we show that the monogamy inequality is satisfied by the GHZ state in arbitrary dimension. In computational basis, the GHZ state can be written as 
\begin{equation}
\ket{GHZ}=\frac{1}{\sqrt d}\sum_{i=0}^{d-1}\ket{iii}.
\end{equation}  
It is well known for GHZ state that if one of the subsystem is traced out, the resulting reduced density matrix becomes a maximally mixed for which $G^d_{ij}$ ($i=1,j=2,3$) vanishes. Again, the Schmidt coefficients when written in the form of Eq. (\ref{schmidt_form}) are obtained $\{\frac{1}{\sqrt d},\frac{1}{\sqrt d},\dots,\frac{1}{\sqrt d}\}$. Thus, we have  $G_{1(23)}=1$ which asserts that a tripartite qudit GHZ state satisfies the monogamy inequality (\ref{ineq}) $G^d_{1(23)}>G^d_{12}+G^d_{13}$. 

\section{Conclusions}\label{discuss}
In this work, we have proved an inequality which expresses a quantitative constraint of sharing entanglement in a pure tripartite qudit state. Our inequality involves G-concurrence measure of entanglement. To establish the monogamy inequality, we have formulated a rigorous framework for estimating an upper bound of G-concurrence of arbitrary dimensional mixed states. It further motivates to obtain a tighter upper bound of G-concurrence that might serve as a quantifier of entanglement of qudit mixed states. It is worth noting that we can reproduce the monogamy inequality akin to CKW inequality \cite{ckw} from our derived monogamy relation (\ref{ineq}). For qubit systems, G-concurrence is realised as concurrence, hence, we trace back the CKW inequality. However, the monogamy inequality based on G-concurrence measure of entanglement is valid for arbitrary dimensional tripartite states. Intuitively, it seems that other concurrence monotones specified in \cite{gour2} would also satisfy certain algebraic constraints, and it is an interesting direction of future study to further explore the monogamous nature of entanglement using these concurrence monotones.   On the other hand, a universal monogamy inequality must involve arbitrary dimensional multipartite systems. Although CKW inequality can be generalised for arbitrary N-partite states, it does not hold for $d>2$. So, it would be worthwhile to investigate whether the inequality (\ref{ineq}) can be extended for the multipartite scenario. It would also be an interesting direction to investigate monogamy exponent \cite{gour4} obtained in the present analysis for G-concurrence measure is the tightest possible bound, or the minimal achievable exponent.
\section*{Acknowledgement}
The author is grateful to Prof. Archan S Majumdar, SNBNCBS, Kolkata, and Prof. Pankaj Agrawal, IOP, Bhubaneswar for providing their valuable insights and suggestions which helped a lot to develop the work. \\~\\

\bigskip 

\onecolumngrid

\appendix 

\section{Proof of the theorem for $d=3$ }\label{appendix1}

Following Eq.(\ref{eq10}), we obtain for $d=3$
\begin{eqnarray}\label{eq18}
G^3_{12}&=&|\lambda_{00}|^2|\lambda_{10}|^2|\lambda_{20}|^2
\end{eqnarray}
 Thus, we rewrite Eq.(\ref{eq18}) as 
 \begin{eqnarray}
G^3_{12}&=&\sum_j|\tilde{a}_{0j0}|^2\sum_j|\tilde{a}_{1j0}|^2\sum_j|\tilde{a}_{2j0}|^2\\
&=&\sum_j|\sum_{k} U_{0k}a_{0jk}|^2\sum_j|\sum_{k}U_{0k}a_{1jk}|^2\sum_j|\sum_{k}U_{0k}a_{2jk}|^2\nonumber\\
\\
&\le & \sum_{k}|U_{0k}|^2\sum_{j,k}|a_{0jk}|^2\sum_{k}|U_{0k}|^2\sum_{j,k}|a_{1jk}|^2\nonumber\\
&&\sum_{k}|U_{0k}|^2\sum_{j,k}|a_{2jk}|^2
\end{eqnarray}
In the last line we have used the Cauchy-Schwarz inequality. 
To evaluate the upper bound of $G^3_{12}$, we recall that $U$ is a unitary matrix, so $\sum_k|U_{0k}|^2=1$. Thus, we obtain  
\begin{eqnarray}
G^3_{12}&\le &  \sum_{j,k}|a_{0jk}|^2\sum_{j,k}|a_{1jk}|^2\sum_{j,k}|a_{2jk}|^2 \label{g-con12}
\end{eqnarray}
In the above, we have found an upper bound of G-concurrence for $\rho_{12}$. 
\\
\\
Now, we proceed to evaluate G-concurrence of the other subsystem $\rho_{13}$ which can be obtained by tracing over the second subsystem in Eq.(\ref{eq4}). As in the previous case, we get the particular pure state decomposition of $\rho_{13}$ by choosing a suitable isometry $V$, and a similar expression can be obtained
\begin{eqnarray}
G^3_{13}&\le & \sum_{k}|V_{0k}|^2\sum_{j,k}|a_{0jk}|^2\sum_{k}|V_{0k}|^2\sum_{j,k}|a_{1jk}|^2\nonumber\\
&&\sum_{k}|V_{0k}|^2\sum_{j,k}|a_{2jk}|^2
\end{eqnarray}

Using $\sum_k|V_{0k}|^2=1$, we find the upper bound of G-concurrence of the subsystem $\rho_{13}$ 
\begin{equation}\label{g-con13}
G^3_{13}\le \sum_{j,k}|a_{0jk}|^2\sum_{j,k}|a_{1jk}|^2\sum_{j,k}|a_{2jk}|^2.
\end{equation}
 Now, adding Eq.(\ref{g-con12}) and Eq.(\ref{g-con13}) we obtain
\begin{eqnarray}
G^3_{12}+G^3_{13}&\le & 2\sum_{j,k}|a_{0jk}|^2\sum_{j,k}|a_{1jk}|^2 \sum_{j,k}|a_{2jk}|^2\nonumber\\
\label{sum12+13}
\end{eqnarray}

For $d=3$, we obtain from Eq.(\ref{gcon_rho_1-23})
\begin{equation}\label{rho1,d=3}
G^3_{1(23)}=3^3\sum_{j,k}|a_{0jk}|^2\sum_{j,k}|a_{1jk}|^2\sum_{j,k}|a_{2jk}|^2
\end{equation}
Comparing Eq.(\ref{sum12+13}) and Eq.(\ref{rho1,d=3}), we obtain
\begin{equation}
G^3_{12}+G^3_{13}\le G^3_{1(23)}.
\end{equation}
This proves the theorem for $d=3$.


\begin{thebibliography}{99}


\bibitem{ckw} Distributed entanglement, V. Coffman, J. Kundu, W. K. Wootters, Phys. Rev. A. \textbf{61}, 052306 (2000).


\bibitem{qkd1} Quantum cryptography based on Bell's theorem, A. K. Ekert, Phys. Rev. Lett. \textbf{67}, 661  (1991).

\bibitem{qkd2} Security proof for cryptographic protocols based only on the monogamy of Bell's inequality violations, M. Paw\l{}owski, Phys. Rev. A \textbf{82}, 032313 (2010).

\bibitem{acin}Asymptotic Quantum Cloning Is State Estimation,  J. Bae and A. Ac\'{i}n, Phys. Rev. Lett. \textbf{97}, 030402 (2006).


\bibitem{ma}Quantum simulation of the wavefunction to probe frustrated Heisenberg spin systems, MaXiao-song Ma, Borivoje Dakic, William Naylor, Anton Zeilinger, Philip Walther \emph{et al.}, Nat. Phys., \textbf{7} 399, (2011).

\bibitem{rao}Multipartite quantum correlations reveal frustration in a quantum Ising spin system, KRK Rao, H Katiyar, TS Mahesh, A Sen, U Sen, A Kumar, Phys. Rev. A, \textbf{88} 022312, (2013).


\bibitem{ben}C. H. Bennett, in Proceedings of the
FQXi 4th International
Conference, Vieques Island,
Puerto Rico, 2014,
http://fqxi.org/conference/talks/2014.

\bibitem{loyd}Unitarity of black hole evaporation in final-state projection models, S. Lloyd and J. Preskill, J. High Energy Phys., \textbf{08} 126, (2014).

\bibitem{con1}Entanglement of a Pair of Quantum Bits, S. Hill and W. K. Wootters, Phys. Rev. Lett. \textbf{78}, 5022 (1997).


\bibitem{con2}Entanglement of Formation of an Arbitrary State of Two Qubits, W. K. Wootters, Phys. Rev. Lett. \textbf{80}, 2245 (1998).

\bibitem{osborne}General Monogamy Inequality for Bipartite Qubit Entanglement, T. J. Osborne and F. Verstraete, Phys. Rev. Lett. \textbf{96}, 220503 (2006).

%


\bibitem{sq}Monogamy of quantum entanglement and other correlations, M. Koashi and A. Winter Phys. Rev.
A, \textbf{69} 022309, (2004).

\bibitem{q}Monogamy of entanglement for tripartite systems, Y. Luo, T. Tian, L.-H. Shao, and Y. Li Phys. Rev. A, \textbf{93} 062340, (2016).








\bibitem{mixed}General monogamy relation for the entanglement of formation in multiqubit systems, Y.K. Bai, Y.F. Xu, and Z. D. Wang Phys. Rev. Lett., \textbf{113}, 100503, (2014).


\bibitem{kim_das}Entanglement monogamy of multipartite higher-dimensional quantum systems using convex-roof extended negativity, Jeong San Kim, Anirban Das, and Barry C. Sanders Phys. Rev. A \textbf{79}, 012329 (2009).


\bibitem{li1}Li, Q., Cui, J., Wang, S. et al. Entanglement monogamy in three qutrit systems. Sci Rep \textbf{7}, 1946 (2017).


 \bibitem{alpha} Monogamy and polygamy for generalized $W$-class states using R\'enyi-$\ensuremath{\alpha}$ entropy, Yanying Liang, Zhu-Jun Zheng, and Chuan-Jie Zhu Phys. Rev. A \textbf{102}, 062428 (2020).



\bibitem{vidal} Entanglement of Pure States for a Single Copy, G. Vidal
Phys. Rev. Lett. \textbf{83}, 1046 (1999).

 
\bibitem{gour2}Family of concurrence monotones and its applications, G. Gour Phys. Rev. A \textbf{71}, 012318 (2005).

\bibitem{sanpera}Schmidt-number witnesses and bound entanglement, A. Sanpera, D. Bru\ss, and M. Lewenstein, Phys. Rev. A \textbf{63}, 050301(R) (2001).




\bibitem{huber}Quantifying Entanglement of Maximal Dimension in Bipartite Mixed States, G. Sent\'{i}s, C. Eltschka, O. G\"{u}hne, M. Huber, and J. Siewert, Phys. Rev. Lett. \textbf{117}, 190502 (2016).
 
\bibitem{yong}Violation of monogamy inequality for higher-dimensional objects, Y.C. Ou, Phys. Rev. A \textbf{75}, 034305 (2007).
 
\bibitem{gour3} Mixed-state entanglement of assistance and the generalized concurrence, G. Gour Phys. Rev. A \textbf{72}, 042318 (2005). 



\bibitem{gour4}Monogamy of entanglement without inequalities, G. Gour and Yu Guo, Quantum \textbf{2}, 81 (2018).

\bibitem{1hjw}Generalized W-class state and its monogamy relation, Jeong San Kim, Barry C. Sanders  	J. Phys. A: Math. Theor. \textbf{41}(49): 495301 (2008).
%



%
%
%
\end{thebibliography}
\end{document}